\documentclass[twocolumn,oneside]{rsauthor}
\setkeys{Gin}{draft=false}
\usepackage[]{graphicx,color,epstopdf}
\usepackage[numbers,sort,sort&compress,comma]{natbib} 
\usepackage{titlesec,amsmath,endnotes}
\usepackage[totalwidth=550pt,totalheight=700pt]{geometry}

\newcommand{\p}{\paragraph}
   
\definecolor{BlueViolet}{RGB}{30, 7, 156} 
\begin{document}

\title{Lag, lock, sync, slip: the many ``phases'' of coupled flagella}
\author{Kirsty Y. Wan, Kyriacos C. Leptos, Raymond E. Goldstein}
\address{Department of Applied Mathematics and Theoretical
Physics, Centre for Mathematical Sciences, University of Cambridge, Wilberforce Road, Cambridge CB3 0WA, UK}
\date{\today}

\abstract{In a multitude of life's processes, cilia and flagella are found indispensable. 
Recently, the biflagellated chlorophyte alga \textit{Chlamydomonas} has become a model organism for the study of ciliary coordination and synchronization.
Here, we use high-speed imaging of single pipette-held cells to quantify the rich dynamics exhibited by their flagella.
Underlying this variabiltiy in behaviour, are biological dissimilarities between the two flagella -- termed \textit{cis} and \textit{trans}, with respect to a unique eyespot.
With emphasis on the wildtype, we use digital tracking with sub-beat-cycle resolution to obtain limit cycles and phases for self-sustained flagellar oscillations.
Characterizing the \textit{phase-synchrony} of a coupled pair, we find that during the canonical swimming breaststroke the \textit{cis} flagellum is consistently \textit{phase-lagged} relative to, whilst remaining robustly \textit{phase-locked} with, the \textit{trans} flagellum.
Transient loss of synchrony, or \textit{phase-slippage}, may be triggered stochastically, in which the \textit{trans} flagellum transitions to a second mode of beating with attenuated beat-envelope and increased frequency. 
Further, exploiting this alga's ability for flagellar regeneration, we mechanically induced removal of one or the other flagellum of the same cell to reveal a striking disparity between the beating of the \textit{cis} vs \textit{trans} flagellum, in isolation. 
This raises further questions regarding the synchronization mechanism of \textit{Chlamydomonas} flagella.
}

\maketitle

\titleformat{\section}
{\color{BlueViolet}\bfseries}
{\color{BlueViolet}\thesection}{1em}{}
\titleformat{\subsection}
{\color{BlueViolet}\bfseries}
{\color{BlueViolet}\thesubsection}{1em}{}
\titleformat{\subsubsection}
{\color{BlueViolet}\bfseries}
{\color{BlueViolet}\thesubsubsection}{1em}{}


\section{Introduction}
Periodicity permeates Nature and its myriad lifeforms. 
Oscillatory motions lie at the heart of many important biological and physiological processes, spanning a vast dynamic 
range of spatial and temporal scales. 
These oscillations seldom occur in isolation; from the pumping of the human heart, to the pulsating electrical signals in 
our nervous systems, from the locomotive gaits of a quadruped, to cell-cycles and circadian clocks,  these different 
oscillators couple to, 
entrain, or are entrained by each other and by their surroundings.  Uncovering the mechanisms and consequences of
these entrainments provides vital insight into biological function.
Often, it is to the aid of quantitative mathematical tools that we must turn for revealing analyses of these intricate 
physiological interactions.

\begin{figure}[t]
	\centering		
	\includegraphics[width=\columnwidth]{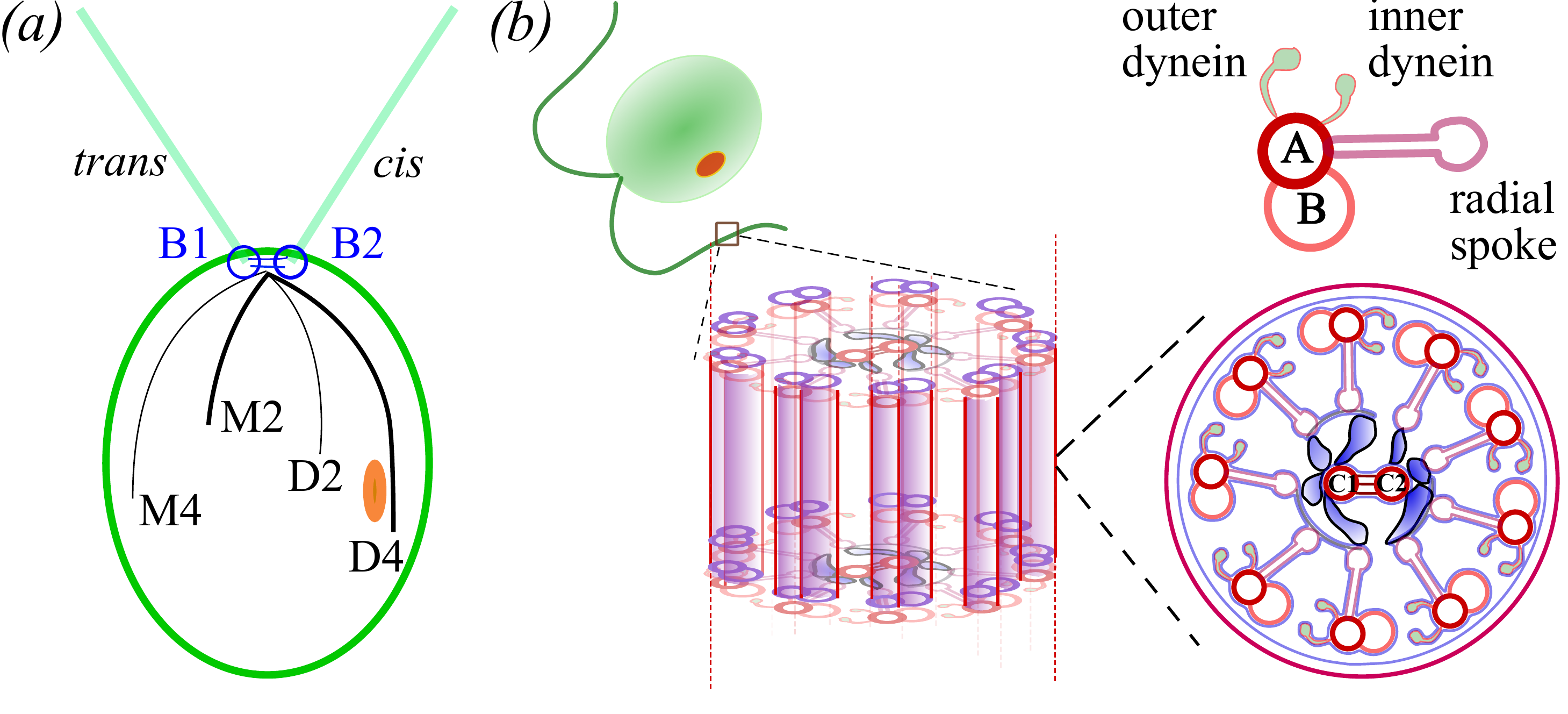}	
	\caption{\small (a) Asymmetric cytoskeletal organization that underlies beating differences between \textit{cis} 
and \textit{trans} flagella. During development, eyespot positioning delineates between the two flagella; 
that closer to the eyespot is the \textit{cis} flagellum, and the farther one is the \textit{trans} flagellum.  
(b) Inside the axoneme: a peripheral arrangement of microtubule doublets encircles a central pair, and specialized 
dynein motors initiate interdoublet sliding and beat generation. }
	\label{fig:axoneme}
\end{figure}

The striking, periodic flagellar beats of one particular organism shall dictate the following discussion: 
\textit{Chlamydomonas reinhardtii} is a unicellular alga whose twin flagella undergo bilateral beating to elicit breaststroke swimming. 
For these micron-sized cells, their motile appendages, termed flagella, are active filaments that are actuated by internal 
molecular motor proteins.
Each full beat cycle comprises a \textit{power stroke} which generates forward propulsion, and a \textit{recovery stroke} 
in which the flagella undergo greater curvature excursions, thereby overcoming reversibility of Stokes flows \cite{Purcell1977}.
A single eyespot breaks cell bilateral symmetry, distinguishing the \textit{cis} flagellum (closer to the eyespot), from 
the \textit{trans} flagellum (figure \ref{fig:axoneme}). 
Subject to internal modulation by the cell, and biochemical fluctuations in the environs, the two flagella undergo a rich variety of tactic behaviours. 
For its ease of cultivation and well-studied genotype, \textit{Chlamydomonas} has become a 
model organism for biological studies of flagella/cilia-related mutations. 
For its simplistic cell-flagella configuration, \textit{Chlamydomonas} has also emerged as an idealised system onto 
which more physical models of flagellar dynamics and synchronization can be mapped 
\cite{Polin2009, Goldstein2009, Geyer2013, Friedrich2013}.
With this versatility in mind, the present article has two goals. 
First, we proffer a detailed exposition of \textit{Chlamydomonas} flagella motion as captured experimentally by
high-speed imaging of live cells; second, we develop a quantitative framework for 
interpreting these complex nonlinear motions. 

In the light of previous work, we have found the motion of \textit{Chlamydomonas} flagella to be sufficiently regular to warrant a low-dimensional phase-reduced description \cite{Polin2009,Goldstein2009,Goldstein2011,Leptos2013}.
Single flagellum limit cycles are derived from real timeseries, and are associated with a phase (\S3.b).
For each cell, dynamics of the flagella pair can thus be formulated in terms of mutually coupled phase 
oscillators (\S3.c), whose pairwise interactions can be determined to 
sub-beat-cycle resolution.
Just as marching soldiers and Olympic hurdlers alike can have preferential 
footedness, we find that \textit{Chlamydomonas} is of no exception; resolving within each cycle of its characteristic breaststroke gait we see that one flagellum is consistently 
phase-lagged with respect to the other.
These transient episodes, previously termed \textit{slips} \cite{Polin2009}, are to be identified with phase slips that 
occur when noisy fluctuations degrade the phase-locked synchronization of two weakly coupled oscillators of differing 
intrinsic frequencies \cite{syncbook}. 
For each cell, sampled here over thousands of breaststroke cycles, and supported by multi-cell statistics, we clarify 
the non-constancy of synchrony observed over a typical cycle.
In particular, the two flagella are found to be most robustly synchronized in the power stroke, and least synchronized 
at the transition to the succeeding recovery stroke.
This trend appears to be universal to all cells of the wildtype strain.
However, the tendency for the two flagella of a given cell to experience phase slips exhibits much greater variation 
across the population (figure \ref{fig:population}).
This we take as further indication that \textit{Chlamydomonas} cells are highly sensitive to fluctuating biochemical cues.
Sampled across multiple cells, phase slip excursions between synchronized states can be visualized by a 
dimension-reduced Poincar\'e return-map of interflagellar phase difference, showing the synchronized state itself to be 
globally stable.
Examining each flagellar phase slip in detail we show further that the \textit{trans} flagellum reproducibly transitions 
to a well-defined transient state with higher beat frequency and attenuated waveform. 
This evidences an alternative mode of beating - which we conjecture to exist as one of a discrete spectrum of modes 
on the eukaryotic flagellum.
This second mode can also be sustained for longer durations, and in \textit{both} flagella of a particular phototaxis 
mutant of \textit{Chlamydomonas}, as detailed elsewhere \cite{Leptos2013}.
Taken together, figure \ref{fig:modes_full} encapsulates in a single diagram the three possible biflagellate ``gaits'', their 
differences and similarities, highlighting the need for a quantitative formulation such as that we present in this article.

Intrinsic differences between the two \textit{Chlamydomonas} flagella, their construction and actuation, underlie this rich assemblage of biflagellate phenomenology.
Past experiments have shown such differences to exist, for example in reactivated cell models of \textit{Chlamydomonas} \cite{Kamiya1984}, in which the \textit{trans} flagellum has a faster beat frequency than the \textit{cis}. 
In contrast, we consider here the \textit{in vivo} case (\S3.d); by mechanically inducing deflagellation of either \textit{trans} or \textit{cis} flagellum, we render live wildtype cells uniflagellate.
This allowed us to compare the intrinsic beating dynamics of \textit{cis} vs \textit{trans} flagella.
We found that whilst \textit{cis}- uniflagellated cells tend to beat with the canonical breastroke-like mode (BS-mode), 
\textit{trans}- uniflagellated cells can instead sustain the faster mode of beating associated with the phase slip (aBS-mode) (figure \ref{fig:CvT_freqs}a). 
Yet this \textit{cis}-\textit{trans} specialization is lost once the cell is allowed to regrow the lost flagellum to full-length, by which time both flagella have recovered the BS-mode.

Flagellar tracking has enabled us to acquire true spatial localization of the flagellum throughout its dynamic rhythmicity,
complementing recent efforts aiming in this direction \cite{Guasto2010,Geyer2013}
The need to know precise waveforms has long been an ambition of historic works, in which manual light-table tracings 
of \textit{Chlamydomonas} flagella were used to elucidate behaviour of the wildtype
\cite{Ruffer1985,Ruffer1987}, and latterly also of flagellar mutants \cite{Ruffer1998}.
We hope that the findings and methodologies herein presented shall be of broad interest to physicists and biologists alike.

\section{Background \label{sec:beat}}

\subsection{The enigmatic flagellum beat}
At a more fundamental level, how is beating of a single flagellum or cilium generated, and moreover, how can 
multi-ciliary arrays spontaneously synchronize?
For each of us, or at least, for the hair-like appendages lining the epithelial cells of our respiratory tracts, this is indeed an important question. 
Beating periodically, synchronously, and moreover metachronously, multitudes of these cilia drive 
extracellular fluid flows which mediate mucociliary clearance. 
These motile cilia, and their non-motile counterparts, are regulated by complex biochemical networks to perform highly-specific functions \cite{Smith2011,Marshall2006}. 
Mutations and defects in these organelles have been increasingly implicated in many human disorders including 
blindness, obesity, cystic kidney and liver disease, hydrocephalus, 
as well as laterality defects such as \textit{situs inversus totalis} \cite{Ibanez-Tallon2003,Fliegauf2007}. 
Mice experiments in which nodal flows are artificially disrupted directly link mechanical flows to positioning of 
morphogens, which in turn trigger laterality-signalling cascades \cite{Nonaka2002}. 

Across the eukaryotic phylogeny these slender, propulsion-generating appendages possess a remarkably conserved ultrastructure \cite{Silflow2001}. 
In recent decades, causality from structure to function within eukaryotic ciliary/flagellar axonemes has been established 
using sophisticated molecular genetics tools.
For the \textit{Chlamydomonas} in particular, rapid freezing of specimens has made possible the capture of axonemal components in near-physiological states, at molecular-level resolution \cite{Nicastro2006}. 
\textit{Chlamydomonas} flagella have a well-characterized $9+2$ structure of microtubule doublets (MTD), along 
which are rows of the molecular motor dynein (figure \ref{fig:axoneme}).
These directional motors generate forces parallel to filamentous MTD tracks, which slide neighbouring MTDs past 
each other.
Anchored to the A-tubule of each MTD by their tail domains, these dyneins detach and reattach onto the next 
B-tubule, consuming ATP in the stepping process.
Different dynein species coexist within the flagellar axoneme, with force-generation principally provided by outer 
dyneins, and modulation of flagellar waveform by the inner dyneins.
The central pair, is thought to transduce signals via the radial spokes \cite{Smith2002}.
Approximately every $96$~nm this precise arrangement of dyneins, radial spokes, and linkers repeats itself \cite{Harris2000}.
Periodic elastic linkages between neighbouring MTDs called nexins provide the coupling by which dynein-driven filament sliding produces macroscopic bending of the flagellum, which in turn propels the cell through the fluid. 
Treatments of axonemes which disrupt dynein domains have shown these nexin linkages to function in a non-Hookean manner to provide the elastic restoring force that resists internal filament shear \cite{Lindemann2005}.
More recently, detailed $3$D-tomographic analysis revealed that nexins are in fact one and the same with the dynein regulatory complex, collectively termed the NDRC \cite{Heuser2009}.
The importance of the NDRC within the functioning axoneme, from those of algae to that of humans, has been highlighted in a recent review \cite{Wirschell2013}.

With regard to the flagellum-cycling mechanism, consensus has been lacking. 
Timing-control appears to be absent, yet much experimental evidence points to selective, periodic, dynein-activation \cite{Nakano2003}.
Both power and recovery strokes of the beat cycle involve active force generation by differentially 
actuated dyneins.
Rhythmic beating of the flagellum may arise from dynamical instabilities of dynein oscillations \cite{Riedel-Kruse2007, Hilfinger2009}, and may be closely coupled to the intrinsic geometry of the axoneme \cite{Brokaw2009, Lindemann2010}.
With these unanswered questions in mind, there is thus much incentive to analyze the flagellum beat \textit{in vivo}, as the present study seeks to demonstrate.
\begin{figure*}[p]

\includegraphics[width=1.0\textwidth]{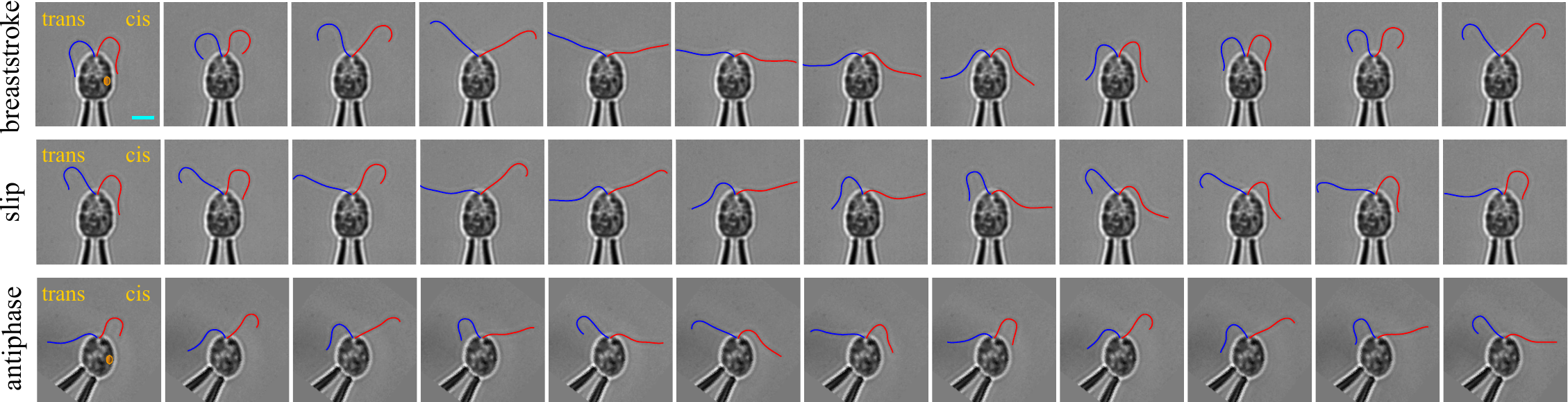} 
\caption{\small Three \textit{Chlamydomonas} breaststroke swimming gaits recorded at $3000$ 
fps and shown at intervals of $5$ frames (i.e. $1.7$ ms intervals). 
Orange dot marks cell eyespot location, red curves: \textit{cis} flagellum, blue curves 
\textit{trans} flagellum. Shown in order, in-phase synchronized breaststroke (both flagella in BS-mode), a phase slip in the same cell (\textit{trans} flagellum in aBS-mode), and 
antiphase synchronization in the phototaxis mutant \textit{ptx1} (both flagella in aBS-mode). Scale bar is $5 \mu m$.}

		\includegraphics[width=0.95\textwidth]{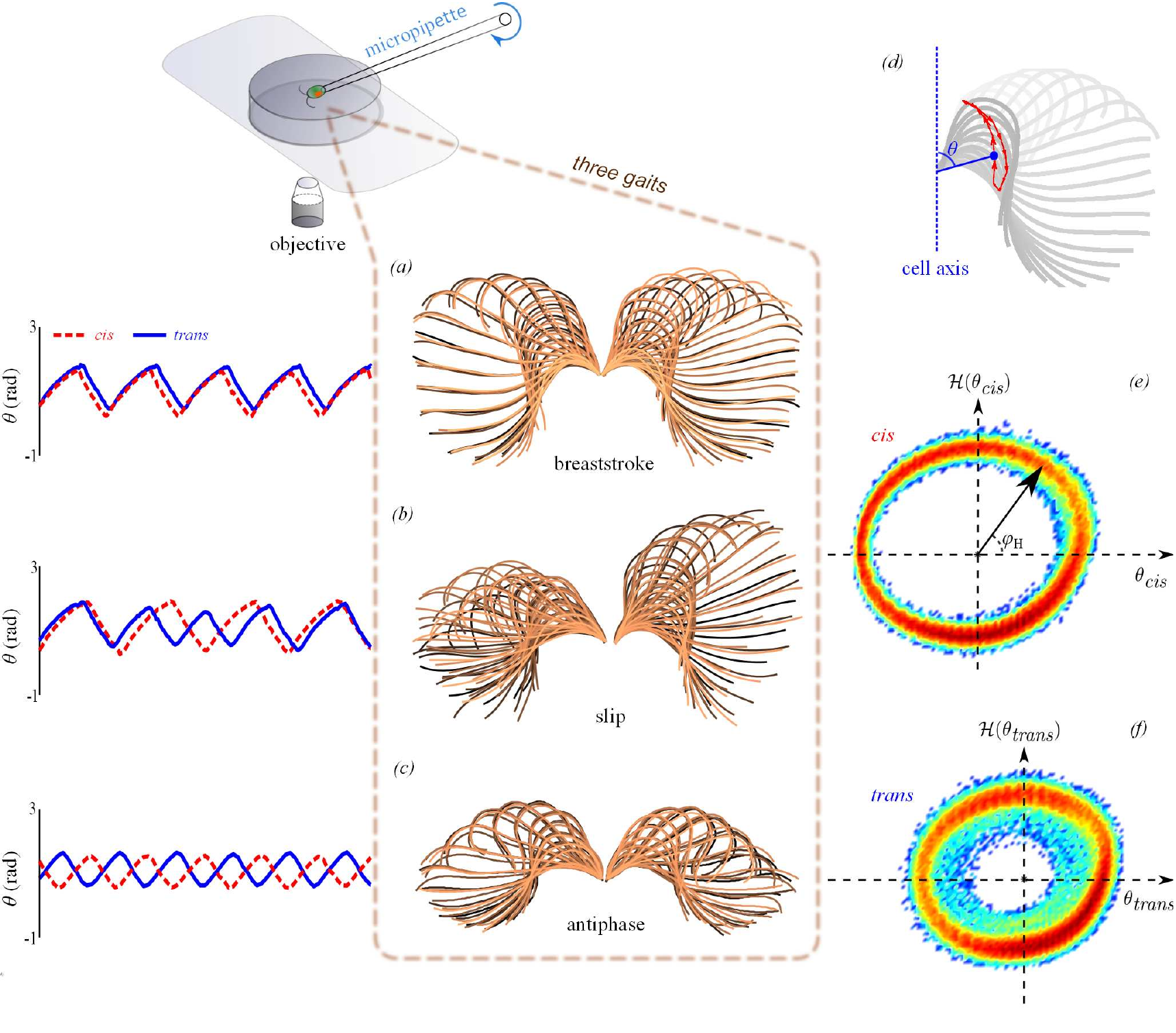}
			\caption{\small Overlaid sequences of tracked flagella showing (a) normal breaststroke (BS) ($5$ consecutive beats), 
(b) a stochastic slip event ($4$ consecutive slips), in which the \textit{trans} flagellum transiently enters a different mode (aBS), 
and (c) in which both flagella of \textit{ptx1} sustain the aBS mode ($5$ consecutive beats).
			For each flagellum, progression through the beat cycle can be tracked using an angle $\theta$ defined relative to the cell bilateral axis (d); sample timeseries for $\theta$ are shown for each gait. 
			The aBS-mode can clearly be seen to have an attenuated beat amplitude and a faster beat frequency. 
			(e)-(f): Differences between \textit{cis} and \textit{trans} limit cycles are clarified in phase-space 
coordinates $(\theta,{\cal H}(\theta))$, where ${\cal H}(.)$ denotes the Hilbert transform (see \S3.b), and the color-intensity is obtained by logarithmically scaling the probability of recurrence. 
	\label{fig:modes_full}}			
\end{figure*}

\subsection{Molecular origins of \textit{cis}-\textit{trans} difference}
As with many species of flagellated algae, motility is essential not just for swimming, 
but also for cell taxis.
For \textit{Chlamydomonas}, its two equal-length flagella, each $10-15$ $\mu$m long, emerge from an 
ellipsoidal cell body $\sim 5$ $\mu$m in radius.
Cell bilateral symmetry is broken by the presence of an eyespot, which functions as a primitive photosensor. 
Perceived directional light is then converted downstream via secondary messenger ions into differential 
flagellar response and hence a turn of the cell \cite{Witman1993,Harris2000,Dieckmann2003}. 
The two anterior basal bodies from which the two flagella protrude are connected to each other via distal striated 
fibres \cite{Ringo1967}.
A system of four acetylated microtubule rootlet-bundles lies beneath the cell membrane, and extends towards the 
cell posterior \cite{Dieckmann2003}.
The eyespot is assembled \textit{de novo} after each cell division, breaking cell bilateral symmetry by its association with the daughter four-membered rootlet (figure \ref{fig:axoneme}a, D4).
The \textit{trans} flagellum, nucleated by the mother basal body (B1), has been shown in reactivated cell models to 
beat at an intrinsic frequency that is $\sim 30-40\%$ higher than that of the \textit{cis} \cite{Kamiya1984}. 
This frequency mismatch is discernible \textit{in vivo} in wildtype cells that we rendered uniflagellated through mechanical deflagellation (\S3.d), additionally with a discrepancy in beating waveform.
Differential phosphorylation of an axonemal docking complex has been suggested to underlie the 
distinctive \textit{cis}-\textit{trans} beat \cite{Takada1997}.
In particular, differential \textit{cis-trans} sensitivity to submicromolar Ca$^{2+}$ in cell models and in isolated axonemes \cite{Kamiya1984, Bessen1980} is consistent with the opposing flagellar modulation necessary for cells to perform phototactic turns \cite{Ruffer1998}.

Yet, despite these intrinsic differences, the two flagella maintain good synchrony during free swimming \cite{Ruffer1985, Guasto2010,Geyer2013}, as well as when held stationary by a micropipette (here). 
Interflagellar coupling may be provided by the motion of the fluid medium \cite{Niedermayer2008,Uchida2011}, by rocking of the cell-body \cite{Friedrich2013}, or further modulated internally via elastic components through physical connections in the basal region \cite{Ringo1967}.
However, stochastically-induced flagellar \textit{phase slips} can appear in otherwise synchronized beating flagella in a distinctive, reproducible manner (figure \ref{fig:modes_full}).
We find that the propensity to undergo these transient slips \cite{Ruffer1998} can vary significantly even between cells of a clonal population (figure \ref{fig:population}).

\section{Results}

\subsection{Three gaits of biflagellate locomotion \label{sec:modes}}
\p{The breaststroke}
In their native habitat, \textit{Chlamydomonas} cells swim in water (kinematic viscosity $\nu=10^{-6}$ m$^{2}$ s$^{-1}$), 
at speeds on the order of $100$ $\mu$m/s, up to a maximum of $200$ $\mu$m/s depending on strain and culture 
growth conditions \cite{Racey1981}.
Oscillatory flows set up by these flagella have a frequency $f=\omega/2\pi\simeq 60$ Hz during the breaststroke.
Stroke lengths of $L=10$ $\mu$m thus produce a tip velocity scale of $U= L\omega \sim 4$ mm/s.
An (oscillatory) Reynolds number $\text{Re}=\omega L^2/\nu$ gauges the viscous and inertial force 
contributions to the resulting flow. 
Here, $\text{Re}\approx 0.001$, and cell propulsion thus relies on viscous resistance to shearing of the fluid 
by the flagellar motion. 
To overcome the reversibility of such flows, a breaking of spatial symmetry is essential during the swimming breaststroke.
The rhythmic sweeping motion of each flagellum can be partitioned into distinct power and recovery strokes: 
during the power stroke, the flagella are extended to maximize interaction with the fluid, but undergo greater bending 
excursions during the recovery stroke (figure \ref{fig:modes_full}a). 
Net swimming progress results from the drag anisotropy of slender filaments and the folding of flagella much 
closer to the cell body during the recovery stroke.
Interestingly, a qualitatively similar bilateral stroke can emerge from a theoretical optimisation performed on swimming 
gaits of biflagellate microorganisms \cite{Tam2011} and on single flagella near surfaces \cite{LaugaPRL}.

\p{The phase slip \label{section:slip}}
Early microscopic analyses of \textit{Chlamydomonas} flagella suggested that the normal breaststroke 
synchrony ``may be disturbed for brief periods'' \cite{Ringo1967}. 
These interruptions to `normal' beating, subsequently detailed in manual waveform tracings by R{\"u}ffer and Nultsch, 
were shown to occur in the absence of obvious stimulation, in both free-swimming cells \cite{Ruffer1985} and cells affixed 
to micropipettes \cite{Ruffer1987}. 
Crucially, these transient asynchronies do not significantly alter the trajectory of swimming (unpublished observation); 
instead, during each such episode the cell body is seen to rock back and forth slightly from frame to frame without altering its prior course. 
These asynchronies are termed \textit{slips} \cite{Polin2009,Goldstein2009} by analogy with an identical phenomenon in weakly-coupled phase oscillators.
Physically, phase slips are manifest in these coupled flagella in a strikingly reproducible manner.
Under our experimental conditions (detailed in \S5), beating of the \textit{trans} flagellum transitions during a slip to a distinct waveform, concurrently with a 
$\sim 30\%$ higher frequency \cite{Leptos2013}, whilst at the same time the \textit{cis}-flagellum maintains its original mode of beating throughout, apparently unaffected  (figure \ref{fig:modes_full}b).
We find also that the faster, attenuated breaststroke mode (aBS) is sustained by the \textit{trans} flagellum for an integer number of full beat cycles, after which normal synchronized breaststroke (BS) resumes. 

\p{The antiphase}
The aBS waveform assumed by the \textit{trans}-flagellum during a slip turns out to be markedly similar to that identified in an anti-synchronous gait displayed by a particular phototaxis mutant of \textit{Chlamydomonas} called \textit{ptx1} \cite{Horst1993}.
In recent, related work, we make these comparisons more concrete, and show that this gait (figure~\ref{fig:modes_full}c) involves actuation of both flagella in aBS-mode, and in precise antiphase with each other \cite{Leptos2013}. 
Although the precise nature of the \textit{ptx1} mutation remains unclear, it is thought that emergence of this novel gait in the mutant is closely associated with loss of calcium-mediated flagellar dominance.


\subsection{Phase-dynamics of a single flagellum \label{sec:single}}
\begin{figure}[t]
	\centering
		\includegraphics[width=0.5\textwidth]{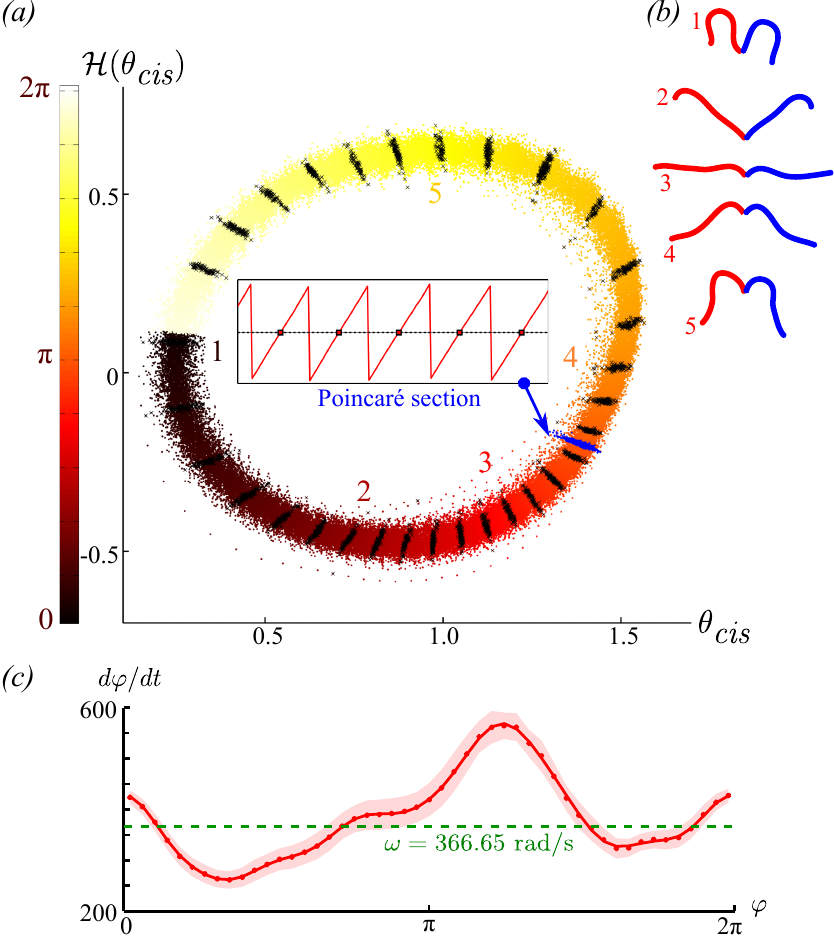}
		\caption{\small (a) Hilbert embedding for the \textit{cis}-flagellum of a sample cell, recorded over thousands of beat cycles, and coloured according to the equivalent transformed phase $\phi$ (via equation~\ref{eqn:phase}). Points of equal phase lie on isochrones, highlighted here at equi-phase intervals. The rate of phase rotation varies systematically throughout the beat cycle, as indicated by the variable inter-isochrone spacing.		
		Inset: successive zero-crossings of $\phi-\left\langle\phi\right\rangle$ map out one Poincar\'e-section. 
		(b) Snapshots $1-5$ show typical positions of the flagellum at phases corresponding to five representative isochrones. 
		(c) Phase-velocity $\Gamma(\varphi)= d\varphi_H/dt$ is approximated by a truncated Fourier series. Shaded regions show one standard deviation of fluctuations in the raw data.  \label{fig:phasetransformation}}
\end{figure}
\begin{figure*}[t]
	\centering
		\includegraphics[]{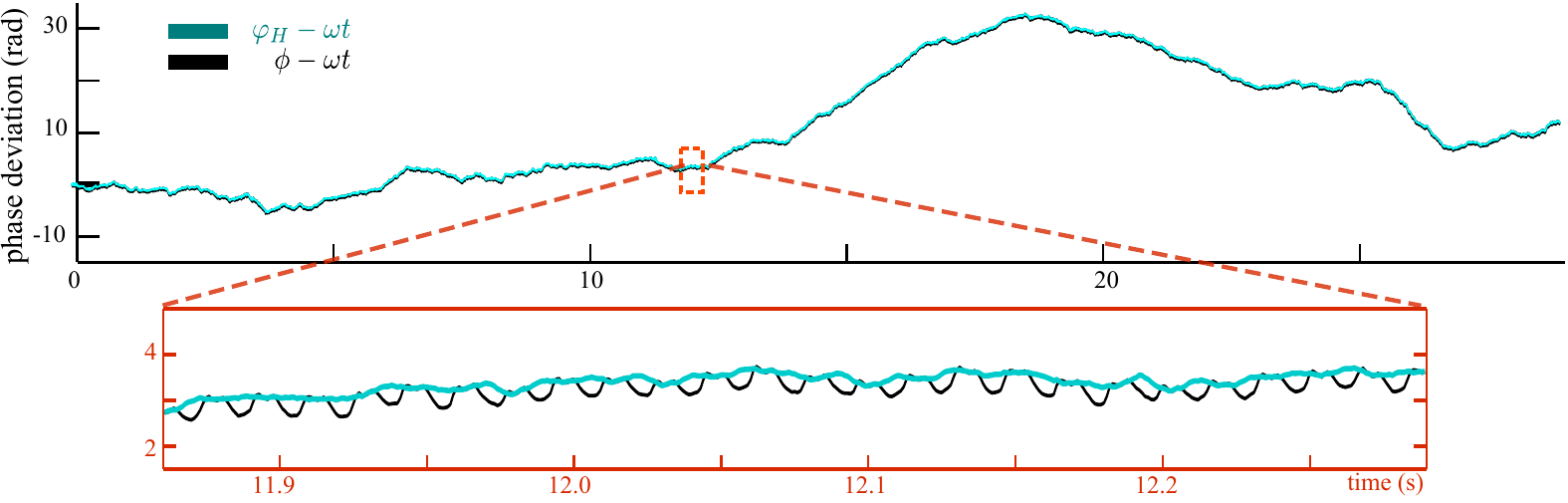}	
    \caption{\small 
		Phase-deviation over long-timescales obtained by subtracting from unwrapped phases the linear component that scales with oscillator frequency. 
		General trends are preserved by the phase transformation (equation~\ref{eqn:gamma}), but within-cycle fluctuations are removed \label{fig:timeseries}}
\end{figure*}

Many biological oscillators are spatially extended and are therefore fundamentally high-dimensional dynamical objects, but adopting a phase-reduction approach facilitates quantitative analyses. 
In such cases, stable self-sustained oscillations can be represented by dynamics on a limit cycle, for which 
monotonically increasing candidate phases $\varphi$ may be extracted.

\begin{figure*}[t]
	\centering
		\includegraphics[width=0.95\textwidth]{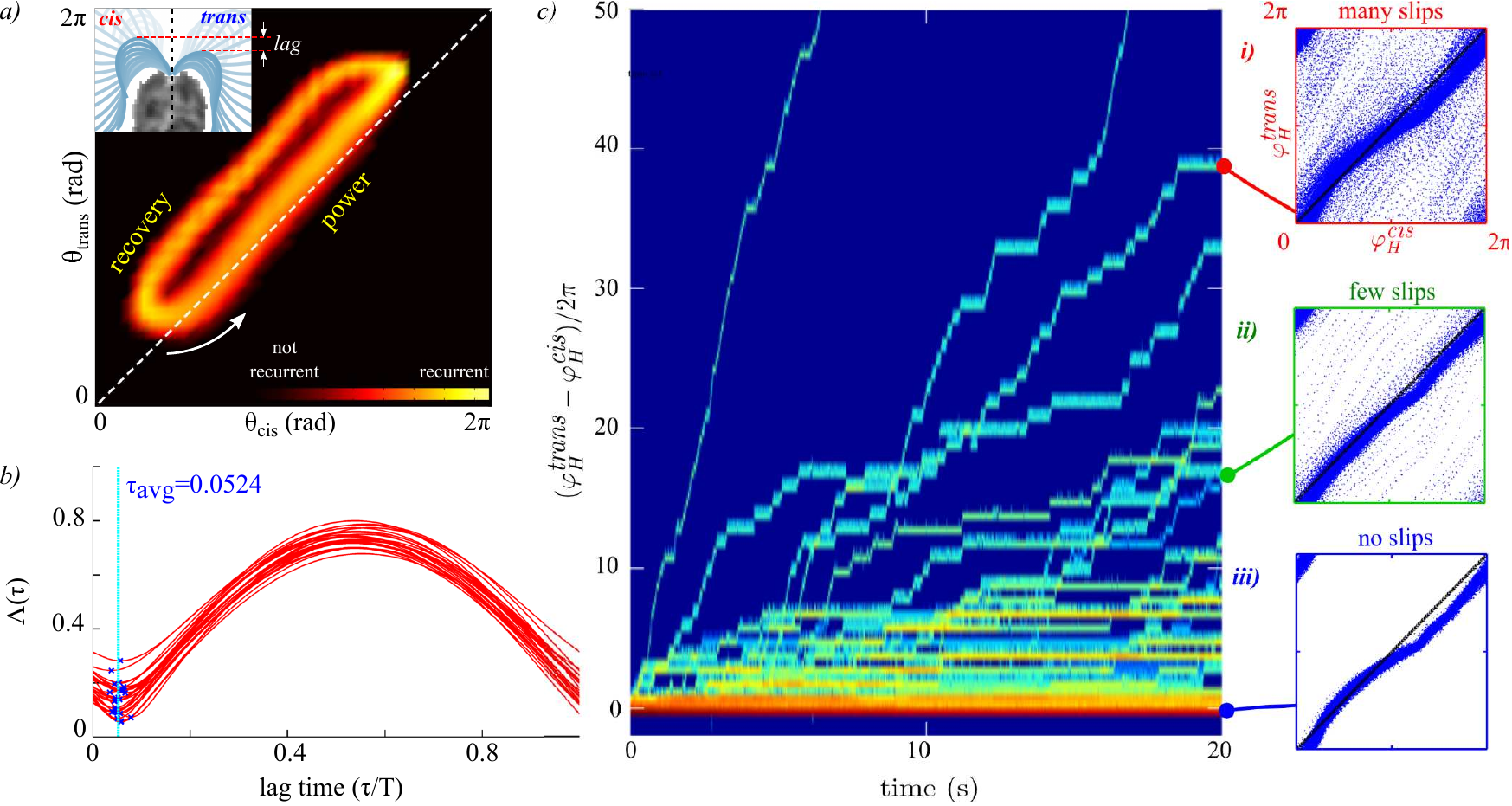}	
	\caption{(a) Lag synchronization in bivariate timeseries of flagella beats, shown for a typical cell. Inset: \textit{cis} flagellum begins its recovery stroke fractionally ahead of \textit{trans}. (b) Multi-cell similarity functions show a similar trend. (c) Noisy flagellar dynamics within a population of $60$ cells, as represented by the tendency of each cell to undergo slips. Most cells remain synchronized for more than $20$~s, whilst some exhibit frequent asynchronies. Red to blue: from high to low probability of recurrence. \label{fig:population}}
\end{figure*}

\begin{figure*}
	\centering
		\includegraphics[width=0.95\textwidth]{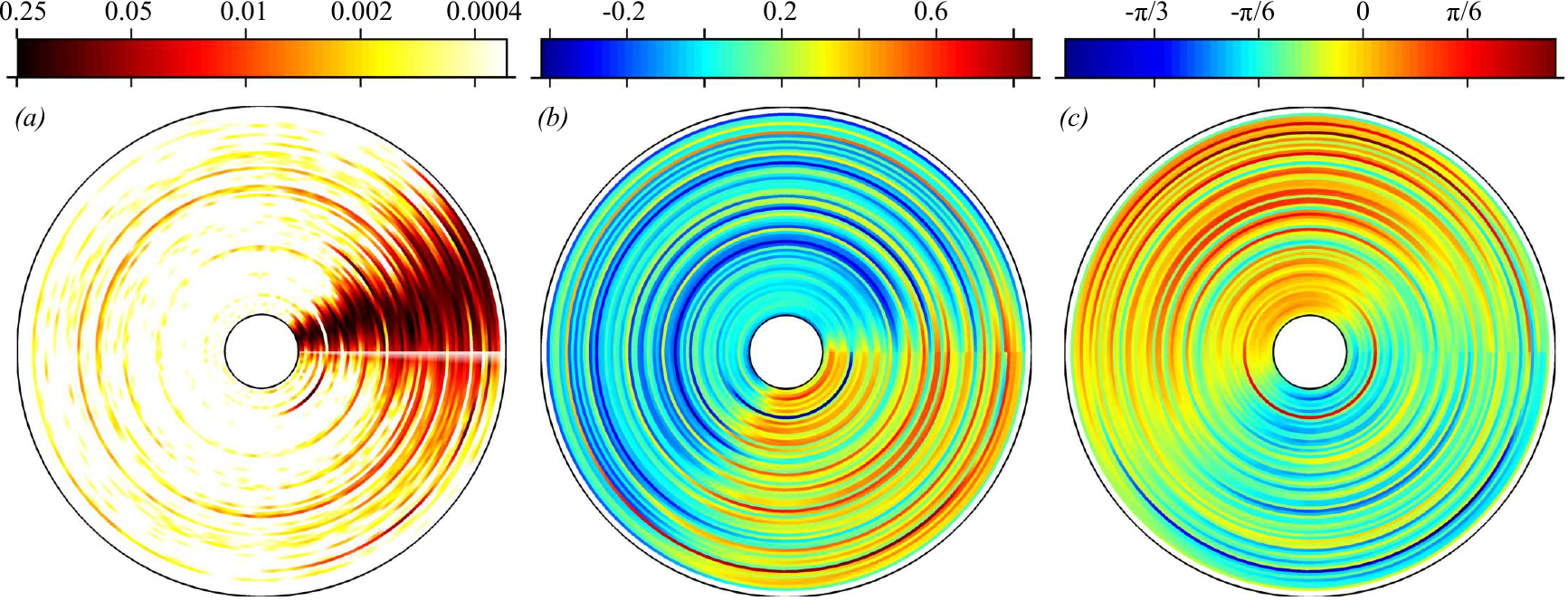}
	\label{fig:fsesync}
	\caption{Trends in \textit{cis}-\textit{trans} flagellar synchronization in a population of \textit{Chlamydomonas} cells: each concentric annulus represents data from an individual cell, measured values are plotted on a circular scale $0\rightarrow2\pi$ in an anticlockwise sense. (a) Probability of stroboscopically observing $\phi^H_{\text{cis}}\in [0,2\pi]$ at discrete marker events where $\theta^H_{\text{trans}}$ reaches a minimum (i.e. start of new power stroke). (b)-(c) Difference in tracked angles ($\Delta_\theta = \theta_{\text{trans}}-\theta_{\text{cis}}$) and Hilbert phases ($\Delta_\phi = \phi^H_{\text{trans}}-\phi^H_{\text{cis}}$), averaged over thousands of beat cycles. It is seen that $\Delta_\theta$ is greatest during the recovery stroke, and correspondingly $\Delta_\phi$ becomes increasingly negative.}	
\end{figure*}

The natural or interpolated phase of an oscillator is defined to increment linearly between successive crossings of a Poincar\'e section, and as such is strongly dependent on the precise choice of the section.
For discrete marker events $\left\{ t_n\right\}$,  and $t_n\leq t\leq t_{n+1}$,
\begin{equation}
\varphi_P(t) = 2\pi\left(n+\frac{t-t_n}{t_{n+1}-t_n}\right)~.
\end{equation}
This method was used in earlier work \cite{Polin2009,Goldstein2009}, to extract flagellar phases $\phi_{\textit{cis},\textit{trans}}$, by sampling pixel intensity variations over pre-defined regions of interest, on individual frames of recorded video. 
Specifically, phase values are interpolated between successive peaks in intensity.
However as beating of the flagellum corresponds to smooth dynamics, by Poincar\'e sectioning the dynamics in this way, sub-beat-cycle information is lost.
In the present work, we take a more continuous approach by incorporating enhanced resolution to elucidate within-beat-cycle dynamics, and make use of flagellum tracking to define $2$D-projections of flagellum oscillations.

For this we choose an embedding via the Hilbert transform. 
This technique derives from the analytic signal concept \cite{Gabor1946}, and is used to unambiguously define an instantaneous phase (and amplitude) from scalar signals with slowly-varying frequency. 
From a periodic scalar timeseries $x(t)$ we construct its complex extension $\zeta(t)=x(t)+i\tilde{x}(t)$, where $\tilde{x}(t)$ is given by
\begin{equation}
{\cal H}(x) = \tilde{x}(t)=\frac{1}{\pi}\text{PV}\int_{-\infty}^{\infty}\frac{x(\tau)}{t-\tau}\,d\tau~.
\end{equation}
Here, the integral to be taken in the sense of the Cauchy principal value.  
Polar angle rotation in the $x-$$\tilde{x}$ phase-plane gives the Hilbert phase,
\begin{equation}
\varphi_H(t) = \tan^{-1}\left(\frac{\tilde{x}-\tilde{x}_0}{x-x_0}\right)~, \label{eqn:Hilbert}
\end{equation}
which, when unwrapped, serves as a monotonically-increasing candidate phase on our limit cycle projection.
The origin $(x_0,\tilde{x}_0)$ is chosen to be strictly interior of the limit cycle; here, $x_0=\left\langle x\right\rangle$, $\tilde{x}_0=\left\langle \tilde{x}\right\rangle$.

Candidate phases such as the Hilbert phase do not uniformly rotate and are sensitive to cycle geometry and nonlinearity.
That is, $\dot{\varphi}=\Gamma(\varphi)$ is in general some non-constant but $2\pi$-periodic function.
For any given limit cycle, there is however a unique true phase $\phi$ for which $\omega=d{\phi}/dt$ is constant, the rate of rotation of $\phi$ is equal to the autonomous oscillator frequency.
The desired transformation $\varphi\rightarrow\phi$ is 
\begin{equation}
\phi = \omega\int^\varphi\,(\Gamma(\varphi))^{-1}\,d\varphi \label{eqn:gamma}~.
\end{equation}
Whilst $\phi$ is unique, ambiguity remains in the choice of limit cycle that best characterizes the original phase-space.
Furthermore for noisy timeseries, limit cycle trajectories do not repeat themselves exactly, so that the phase transformation (\ref{eqn:gamma}) can only be performed in a statistical sense.  
Accuracy of phase estimation is improved with longer observation time.  

To derive an approximation for $d\varphi/dt$ as a function of $\varphi$, we first sort data pairs $\left\langle\varphi,\,d\varphi\right\rangle$ and then average over all ensemble realizations of $\varphi$ (figure \ref{fig:phasetransformation}).
Direct numerical approximations for $dt/d\varphi$ are sensitive to noise, due to the heavy-tailed nature of ratio distributions, to remedy this, we follow the approach of Revzen \textit{et al} \cite{Revzen2008} and begin by finding an N-th order truncated Fourier series approximation $\hat{\Gamma}={\cal F}_N[\Gamma]$, to $\Gamma$.
Next, we find a similar Fourier approximation to $1/\hat{\Gamma}$,

\begin{equation}
  \frac{dt}{d\varphi}(\varphi)\approx{\cal F}_N[1/\hat{\Gamma}]=\sum_{k=-N}^Nf_ke^{ik\varphi}~. \label{eqn:fourier}
\end{equation}
The zeroth coefficient $f_0=\frac{1}{2\pi}\int_0^T\left(\frac{dt}{d\varphi}\right)\,d\varphi=\omega^{-1}$ where $\omega$ is the intrinsic frequency of the oscillator, so that to lowest order, $\varphi$ and $\phi$ coincide.
Substituting ($\ref{eqn:fourier}$) into ($\ref{eqn:gamma}$) gives
\begin{equation}
  \phi = \varphi+2\omega\sum_{k=1}^N\text{Im}\left(\frac{f_k}{k}(e^{ik\varphi}-1)\right)~. \label{eqn:phase}
\end{equation}

For flagellum oscillations, we choose scalar timeseries $x=\theta_{\textit{cis/trans}}$ (Fig.~\ref{fig:modes_full}).
For each flagellum, pairs of values $\mathbf{x}=(x,\tilde{x})$ define a noisy limit cycle in phase space.
Using equation. $\ref{eqn:phase}$, we can associate a phase at each $\mathbf{x}$, consistent with the notion of \textit{asymptotic phase} defined in the attracting region around a limit cycle.
For different flagella, the function $\Gamma(\varphi)$ takes a characteristic form (\ref{fig:phasetransformation}c).
Points of equal phase lie on isochrones, which foliate the attracting domain (figure~\ref{fig:phasetransformation}a).

Perturbations that are $2\pi$ periodic functions of $\varphi$ are eliminated by the transformation \ref{eqn:phase}.
Within-period oscillations are averaged out, whilst long-timescale dynamics are preserved, by virtue of the invertibility of the transformation.
This can be seen over long-time recordings, in which the two measures of phase deviation: $D_\phi=\phi-\omega t$ and $D_\varphi=\varphi_H-\omega t$ coincide (figure \ref{fig:timeseries}), but the periodicity of $\Delta_\varphi$ has been smoothed out (inset).

\subsection{Phase-dynamics of coupled flagella pair\label{sec:pair}}

\begin{figure}[h]
	\centering
		\includegraphics[width=0.85\columnwidth]{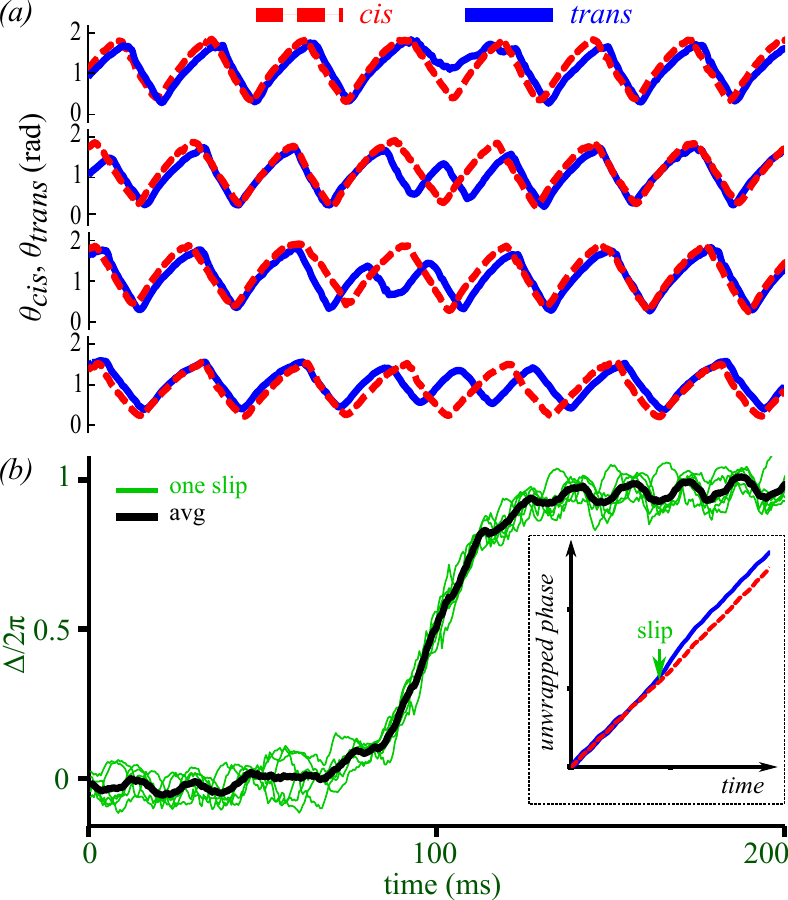}	
	\caption{a) Harmonics of a slip, synchrony resumes after a different (but always integer) number of beats of either flagellum. b) each slip results in a step-like transition in phase-difference $\Delta$. For the same cell, successive slips overlap. Inset: forking of unwrapped phases $\varphi_H^{\textit{cis}}$ and $\varphi_H^{\textit{trans}}$.  \label{fig:slip}}
\end{figure}

For microorganisms that rely on multiple flagella for swimming motility, precision of coordination is essential to elicit high swimming efficacy. 
The bilateral geometric disposition of the two \textit{Chlamydomonas} flagella facilitates extraction of phases for an individual flagellum's oscillations, and in turn, derivation of phase synchrony relations between coupled pairs of flagella.
However, a transformation function similar to equation~\ref{eqn:gamma} that is bivariate in the two phases cannot be derived from observations of the synchronized state alone; therefore in the following, we make use of the Hilbert phase (equation~\ref{eqn:Hilbert}).

\p{Phase difference derivation} 
To monitor biflagellar synchrony, the \textit{phase difference} $\varphi_{\textit{trans}}-\varphi_{\textit{cis}}$ is of particular interest.   
More generically, for coupled noisy phase oscillators $i$ and $j$, the dynamics of each is perturbed by the motion of the other, as well 
as by stochastic contributions:
\begin{equation}
	\dot{\varphi_i}= f(\varphi_i)+\varepsilon_i\,g(\varphi_i,\varphi_j)+\xi_i(t)~,\quad(i=1,2)~,
	\label{eqn:dotphi}
\end{equation}
where $g(\varphi_i,\varphi_j)$ is the coupling function, and $\xi$ is a noise term. 
When $\varepsilon_{1,2}=0$ we recover the intrinsic motion of a single flagellum. 
The two oscillators are said to be $n:m$ \textit{phase-locked} when their cyclic relative phase given by $\Delta_{n,m} = n\varphi_1-m\varphi_2$, satisfies $\Delta_{n,m}=\text{Const.}$.
For noisy or chaotic systems this condition may be relaxed to $\Delta_{n,m}<\text{Const}$.
Here, we define the \textit{trans}-\textit{cis} phase difference from the respective angle signals $\theta_{\textit{cis}}(t)$ and $\theta_{\textit{trans}}(t)$ by
\begin{align}
\Delta &= \varphi_H^{\textit{trans}}(t)-\varphi_H^{\textit{cis}}(t) \\ \nonumber
&= \tan^{-1}\left(\frac{\tilde{\theta}_\textit{trans}(t)\theta_{\textit{cis}}(t)-
\theta_{\textit{trans}}(t)\tilde{\theta}_\textit{cis}(t)}{\theta_{\textit{trans}}(t)\theta_{\textit{cis}}(t)
+\tilde{\theta}_\textit{trans}(t)\tilde{\theta}_\textit{cis}(t)}\right)~,
\end{align}
where $\,\tilde{}\,$ again denotes the Hilbert transform.

We measured $\Delta$ for a large population of cells (Fig. \ref{fig:population}). 
Phase-slip asynchronies are associated with rapid changes in interflagellar phase difference, and appear as step-like transitions that punctuate (sometimes lengthy) epochs of synchronized behaviour for which phase difference is constant. 
We see that over a comparable period of observation time (figure~\ref{fig:population}c:i-iii), pairs of flagella can experience either perfect synchrony, few slips, or many slips \endnote{Of the population of cells analysed for figure \ref{fig:population}c, most were observed under white light.
However a small percentage ($10\%$) were observed under red light, but which for the sake of clarity, have not been explicitly marked out in the figure. 
Whilst in both cases variability in frequency of flagellar slips is observed, we find that on average slips occur more prevalently in cells illuminated by red than by white light (discussed further elsewhere).}.
For the population as a whole, the circular representation of figure \ref{fig:population} facilitates simultaneous visualization of general trends in interflagellar phase-synchrony.

\p{Lag synchronization}
Careful examination of a synchronized epoch shows that $\Delta$ is not strictly constant, but rather fluctuates periodically about a constant value.
During execution of breaststroke swimming, Poincar\'e sectioning of the dynamics has suggested previously that the breaststroke gait is perfectly synchronized \cite{Polin2009,Goldstein2009}.
However, plotting $\theta_\textit{cis}$ against $\theta_\textit{trans}$ (figure~\ref{fig:population}a) we see a consistent lag between the two flagella, which is most pronounced during the recovery stroke.
By computing and minimizing the similarity function
\[
\Lambda(\tau)=\sqrt{\frac{\left\langle(\theta_{\textit{trans}}(t+\tau)-\theta_{\textit{cis}}(t))^2\right\rangle}{\left(\left\langle \theta_{\textit{trans}}^2\right\rangle\left\langle \theta_{\textit{cis}}^2\right\rangle\right)^{1/2}}}
\]
we find this discrepancy to be indicative of lag synchronization.
Here, the periodic angle variables $\theta_\textit{cis/trans}$ are chosen as scalar indicators for the progression of each flagellum through its beat cycle.
In particular, the two phases are synchronized with a time lag $\tau_\text{min} = \min_{\tau}\Lambda(\tau)$, where $\Lambda(\tau)$ assumes a global minimum. 
When the oscillators are perfectly synchronized, $\tau_{\text{min}}=0$.
We calculated $\Lambda(\tau)$ for multiple cells, which displayed a similar profile (figure~\ref{fig:population}b).
With $\tau$ normalized by the average inter-beat period $T$, we see that in every instance the minimum is displaced from $0$ (or equivalently $1$), with an value of $0.0524\pm0.01$, indicative of persistent directional lag.

\p{Stability of fixed points and transients}

A phenomenological model such as equation~\ref{eqn:dotphi} has a convenient dynamical analogy.
Phase difference can be interpreted as particle in a washboard potential $V(\Delta)=-\Delta\delta\omega+\epsilon\cos(\Delta)$, subject to overdamped dynamics $\dot{\Delta}=-dV(\Delta)/d\Delta$. 
Potential minima occur where $\dot{\Delta}=0$, which requires $|\delta\omega/\epsilon|<1$. 
For noise with sufficient magnitude, the particle will have enough energy to overcome the potential barrier, at least transiently. 
Stochastic jumps between neighbouring potential minima, are manifest in coupled flagella as the \textit{slip}-mode (figure~\ref{fig:modes_full}b).

\begin{figure}[t]	
\includegraphics[]{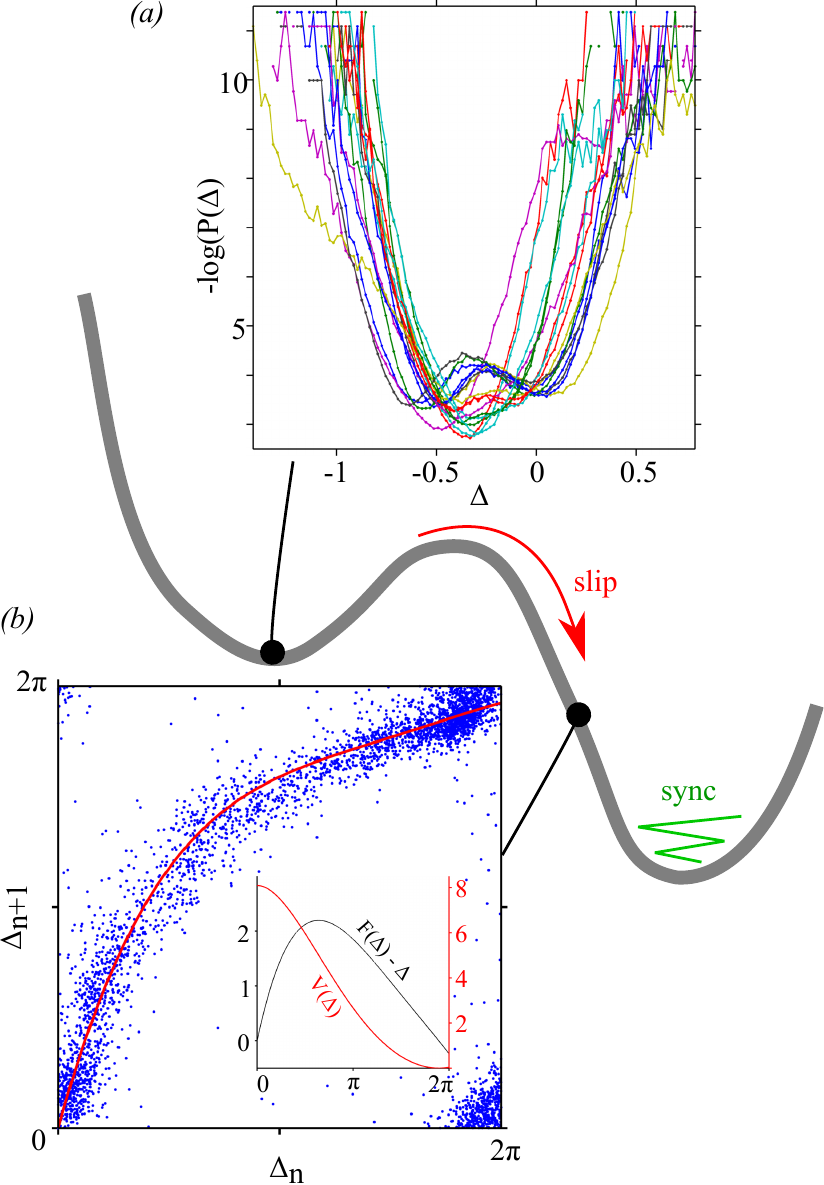}
	\caption{A potential analogy for wrapped phase difference, visualized a) \textit{within} potential minima - $\Delta$ exhibits local fluctuations during the stable synchronized gait; and b) \textit{between} successive potential minima, via a first return map of cyclic/stroboscopic relative phase. 
	An nth-order polynomial $F(\Delta_n)$ is fit to the multicell return-map statistics, from which an empirical potential function $V(\Delta)$ can defined.}\label{fig:wells}
	
	\centering
\end{figure}

In the vicinity of a potential minima, the stationary distribution of $\Delta$ is predicted to be Gaussian (equation~\ref{eqn:dotphi}, with white noise).
This phase distribution $P(\Delta)$, can be measured directly from experiment, and assumed to satisfy
\begin{equation}
P(\Delta)=\exp(-U(\Delta)/k_BT)~,
\end{equation}
from which the potential structure $U(\Delta)$ can be recovered.
For each well, the peak location can be used to estimate the phase-lag of the coupled oscillators, while peak width is indicative of strength of noise in the system.
We measured $P(\Delta)$ for $18$ cells which did not display slips for the duration of observation (figure~\ref{fig:wells}a).
Potential minima have a parabolic profile with a well-defined peak, on average displaced to the left of $\Delta=0$, due to the characteristic phase-lag in the direction of the \textit{cis}-flagellum. 
For certain cells, this lag is especially pronounced during the recovery-stroke than during the power stroke, resulting in a double-peaked minimum in the fine-structure of the empirical potential.

The stability of the synchronized state may be assessed by observing trajectories that deviate from, but eventually return to, this state.
Specifically, by measuring $\Delta$ during multiple flagellar phase slips we construct a dimension-reduced return map for the joint system to visualize the potential landscape that extends between neighbouring minima.
Slips occur with variable duration (figure \ref{fig:slip}a), in which the \textit{trans} flagellum can sustain the faster aBS-mode for a variable but complete number of beat cycles. 
However for an individual cell, successive slips often exhibit identical dynamics (figure~\ref{fig:slip}b).
Figure~\ref{fig:wells}b presents the return map of $\Delta$ associated with $>500$ slip events collected from $70$ cells, where the discretized phase-difference $\Delta_n$ is defined to be $\Delta$ evaluated stroboscopically at the position of maximum angular extent of the \textit{trans}-flagellum, during the $n$th beat-cycle.
We begin by approximating the return map by a polynomial $F(\Delta_n)$.
An empirical potential function \cite{Aoi2013} can be defined by integrating the difference $\delta\Delta=\Delta_{n+1}-\Delta_n\approx F(\Delta_n)-\Delta_n$:
\begin{equation}
\tilde{V}(\Delta_n) = -\int_0^{\Delta_n}\,\delta(\Delta')\,d\Delta'~,
\end{equation}
which we convert into a locally positive-definite function
\begin{equation}
V(\Delta_n) = \tilde{V}(\Delta_n)-\min_{\Delta}\tilde{V}(\Delta)~,
\end{equation}
which satisfies $V(\Delta)>0$, $\forall\,\Delta\notin\text{argmin}\,\tilde{V}(\Delta)$.
The resulting effective potential profile (figure~\ref{fig:wells}b, inset) represents the reproducible phase dynamics of a typical flagellar phase slip, from which breaststroke synchrony re-emerges.  

\subsection{Coupling \textit{cis} and \textit{trans} flagella\label{sec:cistrans} \label{sec:CT}}

Pre-existing, intrinsic differences between the two \textit{Chlamydomonas} flagella are essential for control of cell reorientation, loss or reduction in \textit{cis}-\textit{trans} specialization may give rise to defective phototaxis in certain mutants of \textit{Chlamydomonas} \cite{Okita2005,Leptos2013}.
Under general experimental conditions, stochastic asynchronies which we call slips can punctuate an otherwise synchronous breaststroke; more drastic loss of interflagellar synchrony can lead to \textit{drifts}, which over time, can result in a diffusive random walk in the trajectory of an individual cell \cite{Polin2009}. 
In all these instances, we observe the coupled state of two flagella; in contrast, by mechanically deflagellating wildtype cells (see \S5) we can now examine the intrinsic behaviour of each oscillator in isolation.

\begin{figure*}[t]
	\centering
		\includegraphics[width=0.95\textwidth]{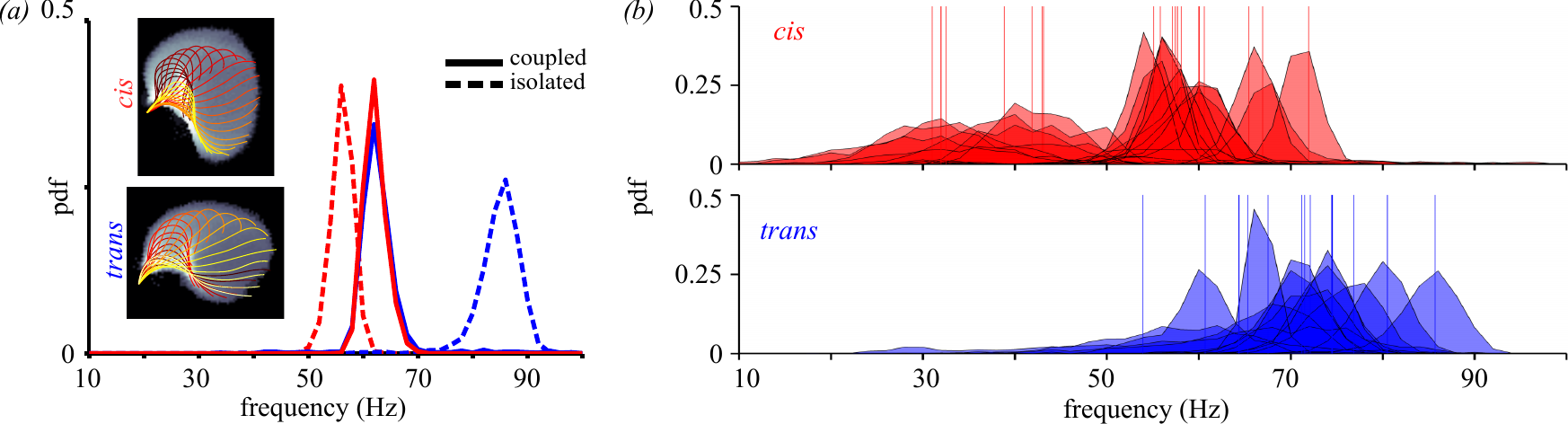}
		\caption{(a) For a single cell, \textit{cis} and \textit{trans} flagella were removed and in turn allowed to regrow to full length. Single-flagellum frequencies separate, but once regrown, lock to a common frequency (\textit{cis}: $57.12$ Hz, \textit{trans}: $80.52$ Hz, \textit{both}: $63.38$ Hz). Insets: typical \textit{cis} and \textit{trans} waveforms, the \textit{trans} waveform is reminiscent of the aBS-mode that onsets during a slip. Waveforms are overlaid on an intensity plot of logarithmically-scaled residence times for each flagellum, over ${\cal O}(100)$ contiguous beats. 
		(b) When beating in isolation, \textit{cis} and \textit{trans} flagella have different frequencies: areas - histograms of interbeat frequencies, lines - averaged frequencies. 
		 \label{fig:CvT_freqs}}
\end{figure*}

The ability of \textit{Chlamydomonas} to readily regenerate a lost flagellum has facilitated controlled measurements of flagellar coupling strength as a function of flagellum length \cite{Goldstein2011}. 
Using the single eyespot as identifier, we removed either the \textit{cis} or \textit{trans} flagellum from a pipette-captured cell and recorded the beating dynamics of the remaining flagellum. 
Histograms of beat-frequencies are plotted in figure~\ref{fig:CvT_freqs}b.
On average, \textit{cis}- uniflagellated cells tend to beat at a lower frequency than \textit{trans}- uniflagellated cells.
A dissociation of beat-frequency of similar magnitude has been observed previously in reactivated cell models \cite{Kamiya1984}. 
Moreover, we find that in the absence of the \textit{cis} flagellum the \textit{trans} flagellum can \textit{sustain} the faster aBS-mode for thousands of beats.
These differences are highlighted in Figure~\ref{fig:CvT_freqs}a, for a single cell.

Interestingly, the aBS-mode that we can now associate with the intrinsic beating waveform of the \textit{trans}, emerges transiently during a slip of the wildtype (figure~\ref{fig:modes_full}b), but in \textit{both} flagella during an antiphase gait of the mutant \textit{ptx1} (figure~\ref{fig:modes_full}c, and \cite{Leptos2013}).
Indeed, for \textit{ptx1}, its lack of effectual \textit{cis}-\textit{trans} specialization has led to speculation that the mutation has renders \textit{both} flagella \textit{trans}-like \cite{Okita2005}.
Specific, structural differences known to exist between the \textit{cis} vs \textit{trans} axonemes \cite{Takada1997} of the wildtype, may effect this segregation of intrinsic beating modes. 

\section{Discussion}
For a unicellular flagellate such as \textit{Chlamydomonas}, synchrony of its two flagella is intimately regulated by the cell's internal biochemistry; however, the exact mechanism by which messenger ions modulate and shape the flagellum beat remains unclear.
Our experimental technique captures the motion of beating flagella \textit{in vivo}, at high resolution, and with respect to a fixed pivot, thereby permitting long-time analysis.

Associating each flagellum oscillator with a continuous phase, we formulated a phase-reduced model of the periodic dynamics.
From long-time series statistics of bivariate oscillator phases, we used phase difference to track phase synchrony, quantifying flagellar interactions for a single individual, as well as across the sampled population.
Exquisitely sensitive to its surroundings, a flagellum can be found to undergo precise, yet dynamic changes when executing its periodic strokes.  
Waveform tracking has allowed us to assess these changes in great spatio-temporal detail. 

In particular, we have found the stable phase-locked breaststroke of \textit{C. reinhardtii} to exhibit a small but persistent \textit{cis}-\textit{trans} phase-lag, the magnitude and direction of which was evaluated from statistics of thousands of beat cycles using a similarity measure, and confirmed for multiple cells. 
However, often it is not the synchronized state itself but rather the emergence or cessation of synchrony that is most insightful for inferring fluctuations in the physiological state of a complex system.
Phase slips are transient excursions from synchrony in which, under our experimental conditions, an alteration of beating mode is observed in the \textit{trans} flagellum only, and that appear to be initiated by a reduced \textit{cis}-\textit{trans} phase lag (figure~\ref{fig:slip}b). 
These reproducible events highlight the importance of \textit{cis}-\textit{trans} specialization of \textit{Chlamydomonas} flagella.
Exploring this further, we mechanically removed individual flagella of wildtype cells to obtain uniflagellated \textit{cis} or \textit{trans} versions, revealing significant differences between their isolated beating behaviours (figure~\ref{fig:CvT_freqs}).
Yet for a fully-intact cell, despite these inherent differences in beating frequency and in waveform, coupling interactions either hydrodynamically through the surrounding fluid medium, and/or biomechanically through elastic linkages at the base of the flagellar protrusion, appear sufficient on the most part to enslave the beating of the \textit{trans} mode to that of the \textit{cis}.

Ours is a very versatile technique for quantifying flagellar synchrony not just of the wildtype system, but such a phase analysis can for instance also be used to probe defective swimming behaviours of motility mutants.
In these cases, macroscopic measurements of population features may not be instructive to understanding or resolving the mutant phenotype, and would benefit from dynamic flagellum waveform tracking and in-depth analysis at the level of an individual cell.

\section{Methods and techniques\label{sec:methods}}

\p{Single algal cells on micropipettes}
For purposes of flagella visualization we chose two wildtype \textit{C. reinhardtii} strains, CC$124$ and CC$125$
(Chlamydomonas Center). 
Stock algae maintained on $2\%$ TAP (Tris-Acetate Phosphate) solid agar slants, were remobilized for 
swimming motility by inoculation into TAP-liquid medium, and cultures used for experimentation were maintained in exponential growth phase ($10^5 - 10^6$ cells/ml) for optimal motility.
Culture flasks are placed onto orbital shakers, and maintained at $24$ $^\circ C$ in growth chambers illuminated on a 
14:10 daily light/dark cycle, so as to imitate the indigenous circadian stimuli. 
Observation of flagellar dynamics was carried out on a Nikon TE2000-U inverted microscope, at constant 
brightfield illumination.
Additional experiments were also performed with a long-pass filter ($622$~nm) to minimize cell phototaxis\endnote{Whilst cell phototaxis behaviour is minimized under red-light illumination, physiological cell motility cannot be maintained in prolonged absence of light, unless the experimental conditions were accordingly modified \cite{Leptos2013}.}.  
Individual cells were captured and held on the end of tapered micropipettes (Sutter Instrument Co. P-$97$), and repositioned with a precision micromanipulator (Scientifica, UK), and imaged at rates of $1,000$-$3,000$ fps (Photron Fastcam, SA$3$). 

\p{Digital Image and Signal Processing}
Recorded movies were transferred to disk for post-processing in \texttt{MATLAB} (Version $8.1.0$, The Mathworks Inc. $2013$.). 
Flagellar waveforms were extracted from individual frames, where contiguous dark pixels that localize the moving flagellum were fit to splines.
Hilbert transforms were perform in \texttt{MATLAB} (Signal Processing Toolbox), and further timeseries analysis performed using custom \texttt{MATLAB} code.

\p{Mechanical deflagellation of either \textit{cis} or \textit{trans} flagella}
To obtain the results described in \S3.d, individual wildtype cells were first examined under white light to locate the unique eyespot, thereby differentiating its \textit{cis} flagellum from the \textit{trans}.
One flagellum was then carefully removed with a second micropipette, by exerting just enough shearing force to induce spontaneous deflagellation by self-scission at the basal region. 
That cells retain the ability for regrowth of flagella ensures basal bodies have not been damaged by our deflagellation treatment. 
Cells for which the beating of the remaining flagellum became abnormal or intermittent, and also for 
which a clear \textit{cis}-\textit{trans} identification could not be made, were duly discarded.

\p{Data}  Examples of high-speed movies, flagellar time series, and other data referenced in this work can be found
at: \url{http://www.damtp.cam.ac.uk/user/gold/datarequests.html}.\\

\small
We thank Marco Polin for discussions. Financial support is acknowledged from the EPSRC, ERC Advanced Investigator 
Grant $247333$, and a Senior Investigator Award from the Wellcome Trust (REG).

\theendnotes

\end{document}